\documentclass[12pt,a4paper]{article}
\usepackage{psfrag,epsf_psfrag,a4local}
\input amssym.def
\input amssym.tex

\begin{document}
\hfill {NSF-ITP-96-150}

\begin{center}
{\LARGE\bf Black hole pairs and \\
\vspace*{0.3cm}

supergravity domain walls}\\
\vspace*{0.7cm}

{\large A. Chamblin}\\
\vspace*{0.1cm}

{\small\it Institute for Theoretical Physics,}\\
{\small\it University of California,}\\
{\small\it Santa Barbara, California 93106--4030, U.S.A.}\\
\vspace*{0.2cm}
{and}\\
\vspace*{0.2cm}

{\large J.$\,$M.$\,$A. Ashbourn-Chamblin}\\
\vspace*{0.1cm}

{\small\it Wolfson College, University of Oxford,}\\
{\small\it Oxford OX2 6UD, England}\\
\end{center}
\vspace*{0.6cm}

{\noindent \small \bf Abstract}\\

{\small We examine the pair creation of black holes in the presence of
supergravity domain walls with broken and unbroken supersymmetry. We 
show that
black holes will be nucleated in the presence of non-extreme, 
repulsive walls which
break the supersymmetry, but that as one allows the parameter 
measuring deviation
from extremality to approach zero the rate of creation will 
be suppressed. In particular,
we show that the probability for creation of black holes in the 
presence of an extreme
domain wall is identically zero, even though an extreme domain 
wall has repulsive 
gravitational energy. This is consistent with the fact that the 
supersymmetric, extreme
domain wall configurations are BPS states and should be stable 
against quantum
corrections. We discuss how these walls arise in string theory, 
and speculate about
what string theory might tell us about these objects.}\\

{\noindent \footnotesize {\it PACS:} 04.70.Dy; 04.60.-m; 11.27.+d; 04.65.+e; 
12.60.J }
\newpage
{\noindent \bf 1. Introduction}\\

{Recently, there has been considerable interest in the study of 
domain walls which
arise in $N = 1$ supergravity (SUGRA) theories. When the dilatonic 
coupling is turned
off, these walls correspond to boundaries between regions of 
isolated vacua of a $N =
1$ supergravity matter potential. Such domain walls and their 
global spacetime
structure are extensively analysed in recent literature ([1], 
[2], [3], [4], [5], [6],
[8]). A common feature of many of these solutions is that the 
walls are repulsive,
i.e., inertial observers perceive the wall as a negative energy 
gravitational source.
Therefore, in analogy with the arguments put forth in [7], we 
expect these SUGRA
configurations to be unstable to quantum tunneling events 
such as black hole pair
production, and indeed this is what we find in this paper.

Since we shall be interested in effects which arise because of the global
gravitational properties of these solutions, we will only sketch 
here the relevant
aspects of the configurations and refer the reader to the literature 
for more detail.

With this in mind, we begin by reiterating that (for now at least) 
we have turned off
the dilaton coupling. It turns out [1] that there are {\it three} 
basic species of
SUGRA domain walls, which can be described as follows:}\\

{\noindent i) {\it Extreme} domain walls, which interpolate between isolated
{supersymmetric} minima.}\\

{\noindent ii) {\it Non-extreme} walls, which are expanding 
`two-centred' bubbles.}\\

{\noindent iii) {\it Ultra-extreme} walls, which are simply false 
vacuum bubbles.}\\

{Within each of these phyla of walls, there are sub-classifications 
which specify more
detailed properties of the solutions.}\\
\newpage
{\noindent \bf Extreme Walls}\\

{First, we describe the different types of {\it extreme} vacuum 
domain walls. By
construction [1], these extreme walls are static, planar and have 
a {\it fixed} energy density}
\[
\sigma = \sigma_{\mbox{\footnotesize ext}}
\]
{which is determined by the values of the cosmological constant on 
each side of the
wall ($\sigma_{\mbox{\footnotesize ext}}$ is the `extremal' value of the 
energy density of the wall,
i.e., $\sigma_{\mbox{\footnotesize ext}}$ is a Bogomol'nyi-type bound which 
extreme walls saturate
[1]). Now, it turns out ([3], [4]) that without loss of generality we 
can take the
extreme wall configurations to be conformally flat, i.e., there 
exists a conformal
factor $A(z)$ (where `$z$' is the normal direction to the wall) 
so that the metric may be written}
\begin{equation}
ds^{2} = A(z)\left(dt^{2} - dz^{2} - dx^{2} - dy^{2}\right)
\end{equation}

{With this ansatz, the extreme vacuum walls can then be classified 
in terms of how the
conformal factor $A(z)$ behaves on each side of the wall. It turns 
out that $A(z)$ has
three basic `signatures', each of which corresponds to a distinct 
`type' of domain wall, as outlined:}\\

{\noindent {\bf Type I:} ~A Type I extreme vacuum wall interpolates 
between a SUSY
anti-de Sitter (adS) vacuum and a Minkowski SUSY vacuum. On the 
Minkowski side ($z < 0$), the conformal factor is constant}
\[
A(z) = 1, ~z < 0
\]
{whereas on the adS side, the conformal factor goes like $\frac{1}{z^{2}}$:}
\[
A(z) \,\longrightarrow\, \frac{3}{|{\Lambda}|z^{2}}, ~z \,\longrightarrow\, 
+\infty
\]
{where $\Lambda$ is the cosmological constant. The surface energy density 
of the wall is then determined to be}
\[
\sigma_{\mbox{\footnotesize ext}} = \frac{1}{4\pi\sqrt{3}} \sqrt{|{\Lambda}|}
\]
{(we are working in units for which $G = c = {\hbar} = 1$). Visually, the 
metric conformal factor looks as is illustrated below:}\\
\vspace*{0.3cm}

\epsfxsize=14cm
\epsfysize=8cm
\psfrag{A(z)}[][]{\raisebox{-1.2cm}{$A(z)$}}
\psfrag{z}[][b]{\raisebox{-0.7cm}{$z$}}
\hspace*{-0.7cm} \epsfbox{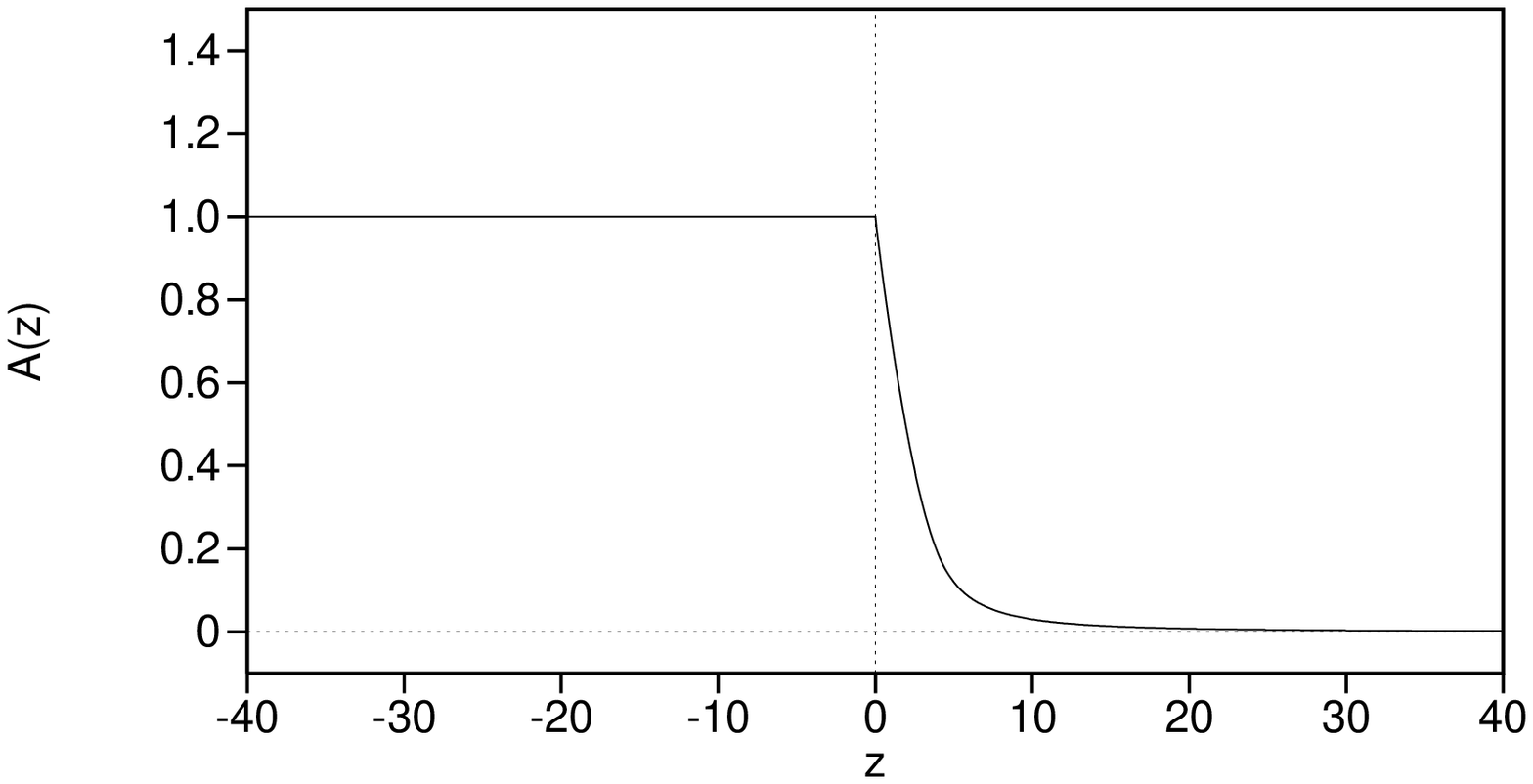}
\vspace*{0.7cm}

{\noindent {\bf Type II:} ~A Type II extreme vacuum wall interpolates 
between two SUSY
adS vacua, such that the conformal factor has the same asymptotic 
behaviour on each
side of the wall,}
\[
A(z) \,\longrightarrow\, \frac{3}{|{\Lambda}_{\pm}|z^{2}} , ~z 
\,\longrightarrow\, \pm
\infty
\]
{where ${\Lambda}_{+}$ is the cosmological constant for $z > 0$ and 
${\Lambda}_{-}$
for $z < 0$ (i.e., ${\Lambda}_{+}$ is not necessarily equal to 
${\Lambda}_{-}$). The
energy density of a Type II wall is calculated to be}
\[
\sigma_{\mbox{\footnotesize ext}} = \frac{1}{4\pi\sqrt{3}}\left( 
\sqrt{|{\Lambda}_{-}|}
\,+\, \sqrt{|{\Lambda}_{+}|}\right)
\]

{\noindent {\bf Type III:} ~A Type III extreme vacuum wall again 
interpolates between
two SUSY adS vacua, only now the conformal factor has different 
asymptotic behaviour
on each side of the wall. Namely, on one side of the wall ($z > 0$, say), 
$A(z)$ falls
off as usual:}
\[
A(z) \,\longrightarrow\, \frac{3}{|{\Lambda}_{+}|z^{2}} , ~z 
\,\longrightarrow\, +\infty
\]
{whereas on the other side $A(z)$ blows up as $z$ approaches some 
critical value $z_{c}$:}
\[
A(z) \,\longrightarrow\, \frac{3}{|{\Lambda}_{-}|(z - Z_{c})^{2}} , ~z
\,\longrightarrow\, z_{c}
\]
{The energy density is then given as}
\[
\sigma_{\mbox{\footnotesize ext}} = \frac{1}{4\pi\sqrt{3}}\left|
\sqrt{|{\Lambda}_{-}|} -
\sqrt{|{\Lambda}_{+}|}\right|
\]
{The point $z_{c}$ is an infinite spacelike (proper) distance away 
from all other points.

In order to analyse geodesic motion in these extreme wall backgrounds, 
the simplest
thing to do is just recover the Lagrangian of motion from the metric 
via the action
principle:}
\begin{equation}
L = A(z)\left({\dot t}^{2} - {\dot z}^{2} - {\dot x}^{2} - {\dot y}^{2}\right)
\end{equation}
{Clearly, no force is felt in either the $x$ or $y$ directions (since 
$A$ only depends
on $z$), and so only the $t$ and $z$ terms yield non-trivial motion. 
Introducing the
energy per mass term $\varepsilon = A{\dot t}$, one integrates 
Equation (2) [10] for
timelike particles ($L = 1$) to obtain the equation of motion}
\begin{equation}
z^{2} - t^{2} = \frac{3}{|{\Lambda}_{\pm}|{\varepsilon}^{2}}
\end{equation}
{Equation (3) with ${\Lambda}_{+}$ applies in the region $z > 0$, 
and likewise for the region $z < 0$.

{}From (3), it is easy to see that on the Minkowski side of a Type I wall, 
freely
falling massive test particles experience no gravitational force. However, 
an inertial
observer on the adS side of a Type I or Type II domain wall is clearly 
moving on a
hyperbola of acceleration {\it away} from the wall. Therefore, in order 
to remain a
fixed distance $z_{0}$ from the wall, an observer would have to turn on 
the rocket
boosters and accelerate {\it towards} the wall. In order to pass through 
the wall from
the adS side, the observer would need an initial velocity of at least 
$\sqrt{1 -
A(z_{0})}$. Observers who start out from $z_{0}$ with less than this 
velocity will be
repulsed.

Thus, both Type I and Type II extremal walls exhibit repulsive behaviour, 
and we would naively expect this repulsive energy to
contribute to the nucleation of black hole
pairs. However, as we shall argue later, the rate of creation in 
these supersymmetric
backgrounds is exactly zero, essentially because the action drops to minus 
infinity very quickly as extremality is approached.
As for Type III walls, although they also have repulsive properties, 
they will
not be considered in this paper, for the simple reason that the side 
of a Type III
wall where the conformal factor diverges is limited by a timelike 
affine boundary.
However, no domain wall-black hole configurations have
such a boundary condition, and so the Type III walls are a priori not
involved with the issue of black hole pair creation.}\\
\vspace*{0.2cm}

{\noindent \bf Non-extreme Walls}\\

{As was pointed out in [1], extreme walls bound SUSY vacua such 
that the gravitational
potentials of the combined vacua {\it exactly} counterbalance the repulsive
gravitational potential of the wall; the result of this exact balance 
is of course the
fact that the wall is actually a static, timelike plane with an 
induced metric {\it
isometric} to $2 + 1$ Minkowski space. When we perturb ${\sigma}$ away from 
$\sigma_{\mbox{\footnotesize ext}}$, and break supersymmetry, we 
expect the wall to
become non-static.

Here, we describe the case where $\sigma > \sigma_{\mbox{\footnotesize 
ext}}$, the
{\it non-extreme} domain walls. Since increasing the energy density 
of the walls
increases the repulsive gravitational energy, we still expect 
these non-extreme walls
to repel inertial observers, and indeed this is what happens. 
In fact, the non-extreme
walls are simply another example of the standard `two-centred' 
bubbles which often
appear as domain wall solutions. The standard vacuum domain 
wall with $\Lambda = 0$ on
each side [12] is such a two-sided bubble; it is known as the 
`Vilenkin-Ipser-Sikivie'
(VIS) solution. In [7] we showed that black hole pairs will 
be nucleated in the
background of VIS, and so it is hardly surprising that a 
similar thing occurs when we
turn on a cosmological constant on each side of the bubble. In fact, a similar
construction was considered in a recent paper of Mann [11], 
although he is still using VIS walls to nucleate the holes 
(i.e., the cosmological
constant is coming from a 3-form field strength).

In more detail, the two-sided bubble is described by taking 
two regions of adS and
glueing them together along a common timelike boundary 
homeomorphic to $S^{2}
\,\times\, {\Bbb R}$. The boundary along which they are 
joined is chosen to satisfy
the Israel matching conditions [12]. Intuitively, to an observer 
on either side of the
bubble, a non-extreme domain wall is a sphere which accelerates 
uniformly away from
them in adS space.

More explicitly, let $\Lambda_{1}$ be the cosmological constant 
on one side of the
wall and $\Lambda_{2}$ the value on the other. Assume that 
on each side of the wall
the spacetime metric assumes the form [1]:}
\[
ds^{2} = e^{2a(z)}\left\{ dt^{2} - dz^{2} - S^{2}(t) \left[ 
(1 - {\chi}r^{2})^{-1}
dr^{2} \,+\, r^{2}d{\phi}^{2}\right]\right\}
\]
{where}
\[ 
S(t) = \left\{ \begin{array}{cl}
1 , &\chi = 0 \\
\cosh \beta t , &\chi = {\beta}^{2}
\end{array} \right. 
\]
{$\beta > 0$ a constant. Then the vacuum Einstein equations reduce to}
\begin{equation}
{\dot a}^{2} - {\beta}^{2} = e^{2a}\left(\frac{-\Lambda}{3}\right)
\end{equation}
{When $\Lambda < 0$ the solution to (4) is}
\[
a(z) = -\ln \left\{ \beta \sinh (\beta z - \beta z^{\prime})
\sqrt{\frac{3}{-\Lambda}}\right\}
\]
{The energy density of such a non-extreme wall is calculated to be}
\[
\sigma = \frac{1}{4\pi}\left( \frac{-\Lambda_{1}}{3} \,+\, 
{\beta}^{2}\right)^{\frac{1}{2}} \,+\,
\frac{1}{4\pi}\left( \frac{-\Lambda_{2}}{3} \,+\, {\beta}^{2}
\right)^{\frac{1}{2}}
\]
{We will begin the next section by considering the pair production 
of black holes in
the presence of these non-extreme walls, since these solutions 
are the most obvious
generalisations of the work described in [7].}\\
\vspace*{0.2cm}

{\noindent \bf Ultra-extreme Walls}\\

{Since ultra-extreme walls have $\sigma < \sigma_{\mbox{\footnotesize 
ext}}$, we again
expect them to be non-static and in fact we expect them to have 
{\it attractive}
gravitational energy. Indeed, this turns out to be the case and so 
these walls cannot
be unstable to tunneling phenomena such as black hole pair production. We will
therefore have nothing more to say about ultra-extreme walls.}\\
\newpage
{\noindent \bf 2. Pair Production of Black Holes by Non-extreme 
Vacuum Walls}\\

{In this section we shall be strictly concerned with the pair 
creation of black holes
carrying a single $U(1)$ charge, i.e., Reissner-Nordstr{\"o}m 
anti-de Sitter (RNadS)
holes. Of course, we could also consider holes which are coupled 
to a dilaton field,
since charged dilaton black holes (with $\Lambda < 0$) which 
are asymptotically adS
are known to exist [13]; however, we shall not consider 
such complications in this
paper. 

Thus, we seek timelike three-surfaces, which satisfy the 
Israel matching conditions,
in the RNadS solution, which is most conveniently written as}
\begin{equation}
ds^{2} = - f(r)\,dt^{2} \,+\, \frac{dr^{2}}{f(r)} \,+\, r^{2}d{\Omega}^{2}
\end{equation}
{where $d{\Omega}^{2}$ is the standard round metric on $S^{2}$ and}
\begin{equation}
f(r) = 1 - \frac{2m}{r} \,+\, \frac{q^{2}}{r^{2}} - \frac{\Lambda}{3} r^{2}
\end{equation}
{Of course, the equation of motion for domain walls in an 
arbitrary (spherically
symmetric) black hole background is well-known [7], [11]. It is given as}
\begin{equation}
\sqrt{f(r) - {\dot r^{2}}} = 2\pi \sigma r
\end{equation}
{where ${\cdot} \,\cong\, f^{-\frac{1}{2}} \partial_{t}$. That 
is, (7) determines the
radial motion of a non-extreme spherical wall surrounding a RNadS black hole.

We are particularly interested in this wall motion on the 
Euclidean section of RNadS.
As in [7], we define the domain wall period, $\beta_W$, 
to be the amount of
`imaginary time' $\tau$ it takes the domain wall to interpolate 
between its minimal
radius, $r_{\mbox{\scriptsize min}}$, and its maximal radius, 
$r_{\mbox{\scriptsize max}}$, on the instanton:}
\[
\beta_W = \oint^{r_{\mbox{\scriptsize max}}}_{r_{\mbox{\scriptsize min}}}\, 
d{\tau} = 
\oint^{r_{\mbox{\scriptsize max}}}_{r_{\mbox{\scriptsize min}}}\, \frac{dr}
{\sqrt{f(f - (2\pi \sigma r)^{2})}}
\]
{i.e., $r_{\mbox{\scriptsize min}}$ and $r_{\mbox{\scriptsize max}}$ 
are the turning 
points where ${\dot r} = 0$.  

As in [7], we now posit the consistency condition, that in order 
for the domain wall
motion to correspond to a well-defined Euclidean section, the wall 
must not intersect
itself. That is to say, the domain wall period ($\beta_W$) must 
be an integer
submultiple of the period of the Euclidean section of RNadS 
(denoted $\beta_{RN}$):}
\begin{equation}
\beta_W = \frac{\beta_{RN}}{n} , ~n \,\in\, {\Bbb Z}_{+}
\end{equation}

{Also, there is the obvious requirement that $r_{min}$ and $r_{max}$ 
both be positive
and real, and this condition leads to a bound on the mass:
$\frac{1}{3\sqrt{3}{\beta}} \leq m \leq \frac{1}{4{\beta}}$.}

{As was pointed out in [14], in order for the RNadS metric (5) to 
describe a charged
black hole in an asymptotically adS space with a non-degenerate 
horizon, the quartic
$r^{2}f(r)$ must have a simple root $r_{0} > 0$ such that $f(r) > 0$ 
for all $r >
r_{0}$. This can occur if and only if the mass satisfies an inequality}
\[
m > m_{c}(q)
\]
{where}
\begin{equation}
m_{c}(q) = \frac{l}{3\sqrt{6}}\left(\sqrt{1 \,+\, 12\left(\frac{q}{l}
\right)^{2}}
\,+\, 2\right)\left(\sqrt{1 \,+\, 12\left(\frac{q}{l}\right)^{2}} -
1\right)^{\frac{1}{2}}
\end{equation}
{where $\Lambda = \frac{-3}{l^{2}}$ as always. Since we desire to 
create holes with
non-degenerate horizons, we will assume this inequality throughout. 
When $q = 0$, $r_{h} = $ `horizon
radius' is defined as the positive solution of $f(r) = 0$. When 
$q \not= 0$, $r_{h}$
is defined as the {\it largest} positive solution of $f(r) = 0$. 
For fixed $q$, then,
$r_{h}$ is a monotone increasing function of $m$:}
\[
r_{c} < r_{h}(m, q) < \infty ~~~\mbox{as}~~~ m_{c} < m < \infty
\]
{where}
\begin{equation}
r_{c}(q) = \frac{l}{\sqrt{6}}\left(\sqrt{1 \,+\, 12\left(\frac{q}{l}
\right)^{2}} -
1\right)^{\frac{1}{2}}
\end{equation}
{The metric is then determined uniquely by $q$ and $r_{0}$; given 
the restriction
$r_{0} > r_{c}$, the mass is totally fixed:}
\[
m = \frac{r_{0}}{2}\left(\frac{r_{0}^{2}}{l^{2}} \,+\, \frac{q^{2}}
{r_{0}^{2}} \,+\,
1\right)
\]

{All properties of the Euclidean section are then functions of 
$r_{0}$ and $q$. In
particular, the period $\beta_{RN}$ is given as}
\begin{equation}
\beta_{RN} = \frac{4\pi r_{0}}{\frac{3r_{0}^{2}}{l^{2}} - 
\frac{q^{2}}{r_{0}^{2}}
\,+\, 1}
\end{equation}

{We now have all the ingredients necessary to apply the 
`no-boundary proposal' to the
calculation of the rate of the black hole pair production in 
these non-extreme SUGRA
domain wall backgrounds. For simplicity, we begin with the 
case where the cosmological
constant is the same on each side of the wall ($\Lambda_{1} = 
\Lambda_{2} = \Lambda$);
we will then make some general comments about the non-symmetric
($\Lambda_{1} \not= \Lambda_{2}$) situation.}\\
\vspace*{0.2cm}

{\noindent \bf Case 1 ($\Lambda_{1} = \Lambda_{2} = \Lambda$)}\\

{Here, we just need to determine the values of $m$ and $q$ 
(or equivalently, $r_{0}$
and $q$, we have fixed $\Lambda$) for which we can satisfy 
the following two relations simultaneously:}
\[
[C] ~~\left\{ \begin{array}{c}
m > m_{c}(q) \\
 \\
\beta_W = \frac{\beta_{RN}}{n} , ~n \,\in\, {\Bbb Z}_{+} 
\end{array} \right.
\]

{The line integral for $\beta_W$ must be performed numerically. Below, we plot
the two periods $\beta_W$ and $\beta_{RN}$, as functions of charge $q$;
the range of q runs from zero (where we recover Schwarzschild-adS black holes)
to an upper bound, $q_{max}$, determined by the non-degeneracy condition 
$m > m_{c}(q)$: Fig. 2}\\
\vspace*{0.3cm}

\epsfxsize=14cm
\epsfysize=8cm
\psfrag{B_W}{$\beta_W$}
\psfrag{B_RN}{$\beta_{RN}$}
\psfrag{B_W and B_RN}[][]{\raisebox{-1.2cm}{$\beta_W$ and $\beta_{RN}$}}
\psfrag{log10(1 - q/qmax)}[][b]{\raisebox{-0.7cm}{$\log_{10}(1 - 
q/q_{\mbox{\scriptsize max}})$}}
\hspace*{-0.7cm} \epsfbox{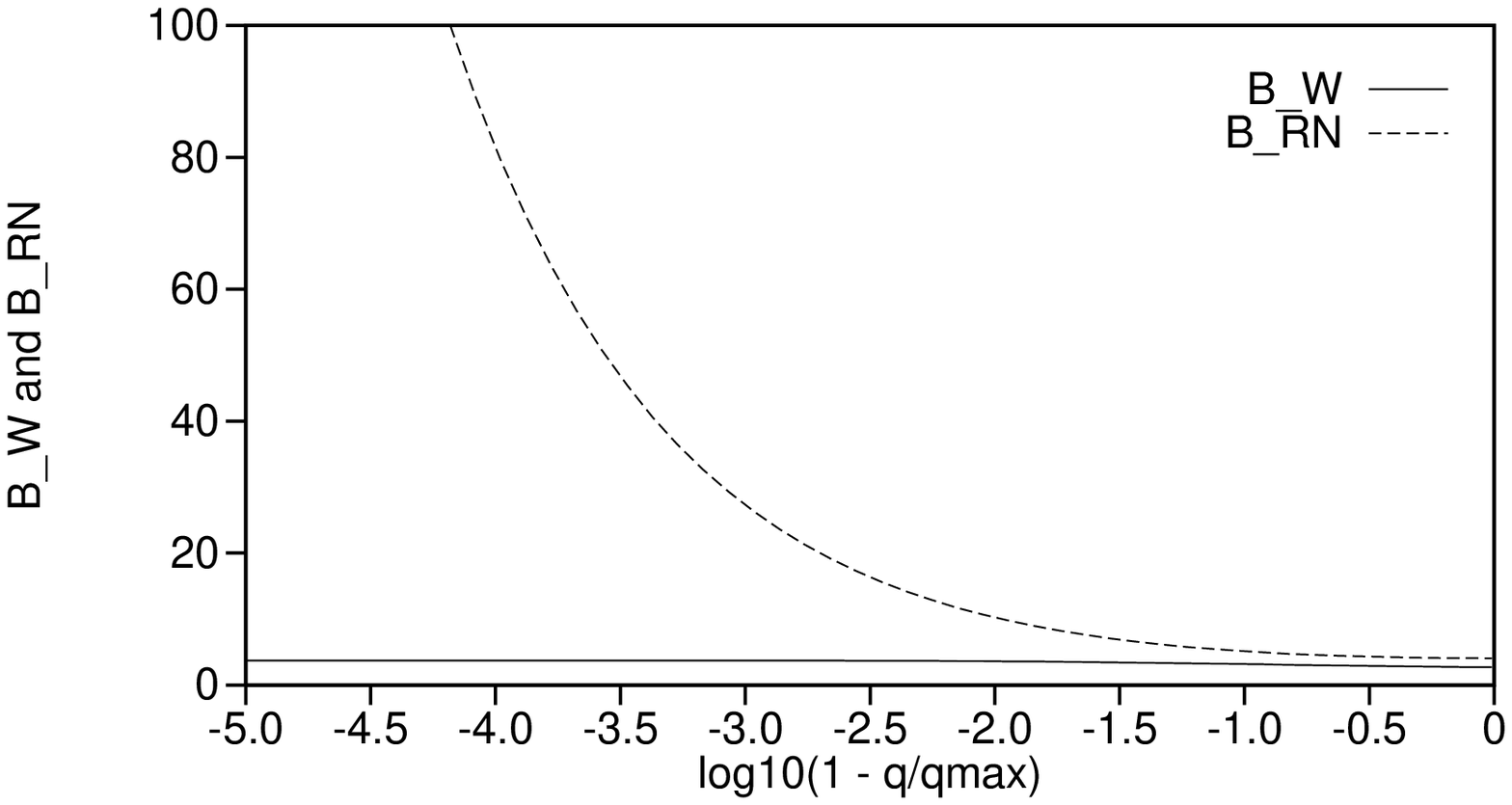}
\vspace*{0.7cm}

{What this figure tells us is clear: For given values of the 
energy density of the
wall (here, $\sigma = \frac{1}{2\pi}\sqrt{\frac{|\Lambda|}{3} 
\,+\, \beta^{2}}$) 
and the mass $m$,
there exists a countably infinite set of charges $\{q_{n}\}$, 
for which it is possible
to find instantons satisfying the constraints $[C]$ (above) 
which will mediate the
creation of accelerating black hole pairs from the 
${\Bbb Z}_{2}$-symmetric initial
domain wall state.

Finally, we point out that as usual there exists a 
`static limit' domain, i.e., ${\dot r} = 0$. 
This static wall lies at $r_{s}$, 
which is calculated to be $r_{s} = {3 \over 2} m \Big[ 1 +\sqrt{1 - 
{8 \over 9} {q^2 \over m^2}} \Big]$

As usual, we calculate the rate of black hole pair creation by 
dividing the amplitude
to create the combined black hole-domain wall configuration by 
the amplitude to create
just the domain wall configuration. In each case, the amplitude is given as}
\[
{\cal A} = e^{- {\cal S}}
\]
{where ${\cal S}$ is half the action of the Euclidean section 
of the relevant solution.

In general, the action in Einstein-Maxwell theory with a negative cosmological
constant and a domain wall boundary term is written as}
\begin{equation}
{\cal S} = \frac{1}{16\pi} \int_{M_{e}} d^{4}x \sqrt{g} \left(R - 2\Lambda -
F^{2}\right) + \frac{1}{2} \sigma \int_W \sqrt{h} \,d^{3}x
\end{equation}
{where $R$ is the four-dimensional Ricci scalar, $\sigma$ is 
the energy density of the
wall and}
\[
\int_W \sqrt{h} \,d^{3}x = \mbox{vol}(W)
\]
{is the volume of the closed, three-dimensional surface 
determined by the domain wall
on the Euclidean section.

We begin by calculating the action of the Euclidean 
section of the non-extreme,
$\Lambda_{1} = \Lambda_{2} = \Lambda$ domain wall solution. Here,}
\begin{eqnarray}
\sigma = \frac{1}{2\pi}\sqrt{\frac{|\Lambda|}{3} \,+\, \beta^{2}} \nonumber \\
F^{2} = 0
\end{eqnarray}
{Since each side of the wall is a portion of adS,}
\begin{equation}
R = 4\Lambda
\end{equation}
{The Euclidean section, $M_{e}$, is obtained by glueing two 
hyperbolic four-balls
together along their boundary three-spheres; one then sees the $M_{e}$ is
topologically $S^{4}$, only with a `ridge' of curvature 
along the domain wall $W
= S^{3}$. Thus, it simply remains to determine $\mbox{vol}(M_{e})$ and 
$\mbox{vol}(W)$.

To calculate these quantities, first recall [15] that 
the volume of a {\it sphere} of
geodesic radius $r$ in hyperbolic space of constant 
negative sectional curvature
$\kappa$, is given as}
\begin{equation}
\mbox{vol}(S^{3}(r)) = 2\pi^{2} \left(\frac{1}{\sqrt{|\kappa|}}\, \sinh
\left(\sqrt{|\kappa|r}\right)\right)^{3}
\end{equation}
{Now, on a four-manifold $R = 4\kappa$, and so $\Lambda 
= \kappa$ in our case; also,
in the above we are working in `exponential' coordinates, so that $r$ is the
coordinate obtained by projection of the usual radial 
coordinate on ${\Bbb R}^{4}$
(recall that the map $\exp:\,{\Bbb R}^{4} \,\longrightarrow\, 
H_{\kappa}^{4}$, where
$H_{\kappa}^{4}$ is hyperbolic space, is in fact a global 
diffeomorphism since the
injectivity radius of $H_{\kappa}^{4}$ is infinity).

Given Equation (15) for the volume of a three-sphere of radius $r$ in
$H_{\Lambda}^{4}$, it is then obvious that the volume of a ball 
of radius $r$ is given
as}
\begin{equation}
\mbox{vol}(B^{4}(r)) = 2\pi^{2} \int_{0}^{r}\left(\frac{1}
{\sqrt{|\Lambda|}}\, \sinh
\left(\sqrt{|\Lambda|r}\right)\right)^{3} dr
\end{equation}
{Since $M_{e}$ is obtained by glueing two of these balls 
together, we need only ask:
what is the radius, $r$?

The radius of the balls is determined by the matching condition 
for joining the
instanton to the Lorentzian section; in particular, we {\it must} 
match the Euclidean
and Lorentzian sections along a spacelike three-surface of vanishing extrinsic
curvature (so that we can think of the tunneling process in 
terms of a `path' of
spacelike three-geometrics). In other words, we match the 
two sections along the
surface where the wall is (instantaneously) stationary. 
Since the wall is simply
moving with uniform acceleration in adS, this surface 
is not too difficult to locate.
The most transparent way to do the calculation is to 
map the adS coordinates used
above to the {\it Einstein cylinder}~coordinates 
(recall that adS is conformalm to
`one half' the Einstein static universe [17]). In terms of 
the coordinates $(t, z, r,
\phi)$, the new coordinates are}
\begin{eqnarray}
T = \xi \sinh(\beta t) \nonumber \\
R = \xi \cosh(\beta t)
\end{eqnarray}
{where $\ln \xi = \beta(z - z^{\prime})$. The domain wall 
(at $z = 0$) then lives on
the hyperbolic trajectory [16]}
\begin{equation}
R^{2} - T^{2} = \frac{1}{\delta}
\end{equation}
{where}
\[
\delta = \frac{\frac{|\Lambda|}{3} \,+\, 2\beta^{2} \,+\, 
2\beta\left(\beta^{2} \,+\,
\frac{|\Lambda|}{3}\right)^{\frac{1}{2}}}{\frac{|\Lambda|}{3}}
\]
{The wall is stationary at $T = t = 0$, and so the `critical' 
value $(R_{c})$ of the
radius is given by}
\begin{equation}
R_{c} = \frac{1}{\beta}
\end{equation}

{Where this is the radius measured in the metric of hyperbolic space, i.e., 
the metric induced by the Gaussian normal coordinates, 
and so $R_{c}$ is the actual
geodesic radius of the instanton.

The volume calculation is time-consuming and yields}
\begin{eqnarray}
\frac{1}{2} \mbox{vol}(M_{e}) = 2\pi^{2} \int_{0}^{R_{c}}
\left(\frac{1}{\sqrt{|\Lambda|}}\,\sinh\left(\sqrt{|\Lambda| r}
\right)\right)^{3} dr
\nonumber \\
= \frac{4\pi^{2}}{\left(\sqrt{|\Lambda|}\right)^{5}}\left\{ \frac{3}{4}\sinh
\sqrt{|\Lambda|R_{c}} - \frac{3}{4}\sqrt{|\Lambda|R_{c}}\,
\cosh\sqrt{|\Lambda|R_{c}}
\right. \nonumber \\
\left. - \frac{1}{36}\sinh\left(3\sqrt{|\Lambda|R_{c}}\right) +
\frac{\sqrt{|\Lambda|R_{c}}}{12}\,\cosh\left(3\sqrt{|
\Lambda|R_{c}}\right)\right\} 
\end{eqnarray}

{The volume of the wall is just the volume of a three-sphere 
of radius $R_{c}$:}
\begin{equation}
\mbox{vol}(W) =
\frac{2\pi^{2}}{\left(\sqrt{|\Lambda|}\right)^{3}}\,\sinh^{3}\left(
\sqrt{|\Lambda|R_{c}}\right)
\end{equation}
{Thus the total action, ${\cal S}_{W}$, for the Euclidean 
section of the ${\Bbb
Z}_{2}$-symmetric non-extreme domain configuration is}
\begin{eqnarray}
\frac{1}{2}{\cal S}_{W} = {\frac{\Lambda}{8 \pi}}\mbox{vol}(M_{e}) +
\frac{\pi^{2}\sigma}{2\left(\sqrt{|\Lambda|}\right)^{3}}\,\sinh^{3}\left(
\sqrt{|\Lambda|R_{c}}\right)
\end{eqnarray}

{We now just need to calculate the action, ${\cal S}_{RN}$, 
of the Euclidean section
of the ${\Bbb Z}_{2}$-symmetric RNadS-domain wall configuration. 
We have to deal with
the extra $F^{2}$ term. As usual, we can either consider 
electrically charged holes,
for which}
\[
F = -\frac{q}{r^{2}}\,dt \,\wedge\, dr
\]
{or magnetically charged holes, for which}
\[
F = q\sin\theta\,d\theta \,\wedge\, d\varphi
\]

{In this paper we shall restrict our attention to the 
creation of magnetically charged
holes; by the usual duality arguments [18] we expect the rate of electric hole
production to be the same (ignoring the 1-loop effects 
of any matter fields, which
obviously would violate electric-magnetic duality).

We begin by calculating the probability to create 
static holes, i.e., holes whose
attractive gravitational energy exactly counterbalances 
the repulsive energy of the
wall. Thus, as we saw above, the black holes will lie 
at radius $r_{s}$ from the wall,
where}
\begin{equation}
r_{s} = {3 \over 2} m \left[ 1 +\sqrt{1 - {8 \over 9} {q^2 \over m^2}} \right]
\end{equation}

{The mass, $m$, is
completely determined in terms of $q$ and $\Lambda$ 
(using relations (10)--(11)
above). Now, as discussed above, in order for the motion of the wall on the 
black hole instanton to be consistent, the
mass $m$ is further constrained to lie in the interval 
$\frac{1}{3\sqrt{3}{\beta}} \leq m \leq \frac{1}{4{\beta}}$. 
Thus, as the acceleration
parameter ${\beta}$ is increased, the mass is forced to 
decrease (as is $r_{s}$, in light
of Equation 23).

The action calculation is now much more straightforward. Let $r_{h}$ denote
the outer horizon radius as above. Then the volume contribution is}
\begin{equation}
\frac{1}{16\pi} \int_{B_{e}} \left(2\Lambda - F^{2}\right) \sqrt{g} d^{4}x = 
\beta_{RN}\Lambda \left[ \frac{r_{s}^{3} - r_{h}^{3}}{12}\right] \,+\, 
\frac{q^{2}}{4}
\beta_{RN} \left( \frac{1}{r_{h}} - \frac{1}{r_{s}} \right)
\end{equation}
{where $B_{e} \,\cong\, D^{2} \,\times\, S^{2}$ is one-half of the Euclidean
section of the black hole-domain wall configuration.

The domain wall term is}
\begin{equation}
 \sigma \int_{W} \sqrt{h} d^{3}x = 4\pi\sigma r_{s}^{2} \beta_{RN} 
\sqrt{f(r_{s})} 
\end{equation}

{Thus, the total action ${\cal S}_{RN}$ is given as}
\begin{equation}
\frac{1}{2}{\cal S}_{RN} = \beta_{RN}\Lambda \left[ \frac{r_{s}^{3} -
r_{h}^{3}}{12}\right] \,+\, \frac{q^{2}}{4} \beta_{RN} \left( \frac{1}{r_{h}}
- \frac{1}{r_{s}}\right) + 4\pi\sigma r_{s}^{2} \beta_{RN} \sqrt{f(r_{s})} 
\end{equation}
{The probability, $P$, that static black holes of charge $\pm q$ and mass $m$
will be created in the presence of the ${\Bbb Z}_{2}$-symmetric wall
background is then given as}
\begin{equation}
P = \exp \left( {\cal S}_{W} - {\cal S}_{RN}\right)
\end{equation}
{where ${\cal S}_{W}$ and ${\cal S}_{RN}$ are given in equations (22)
and (26) respectively.

$P$ is plotted below, for fixed $\Lambda$, $m$, and $\beta$ (and hence
$\sigma$) as $q \,\longrightarrow\, q_{\mbox{max}}$: Fig.3}\\
\vspace*{0.3cm}

\epsfxsize=14cm
\epsfysize=8cm
\psfrag{P}[][]{\raisebox{-0.8cm}{$P$}}
\psfrag{log10(1 - q/qmax)}[][b]{\raisebox{-0.7cm}{$\log_{10}(1 - 
q/q_{\mbox{\scriptsize max}})$}}
\hspace*{-0.7cm} \epsfbox{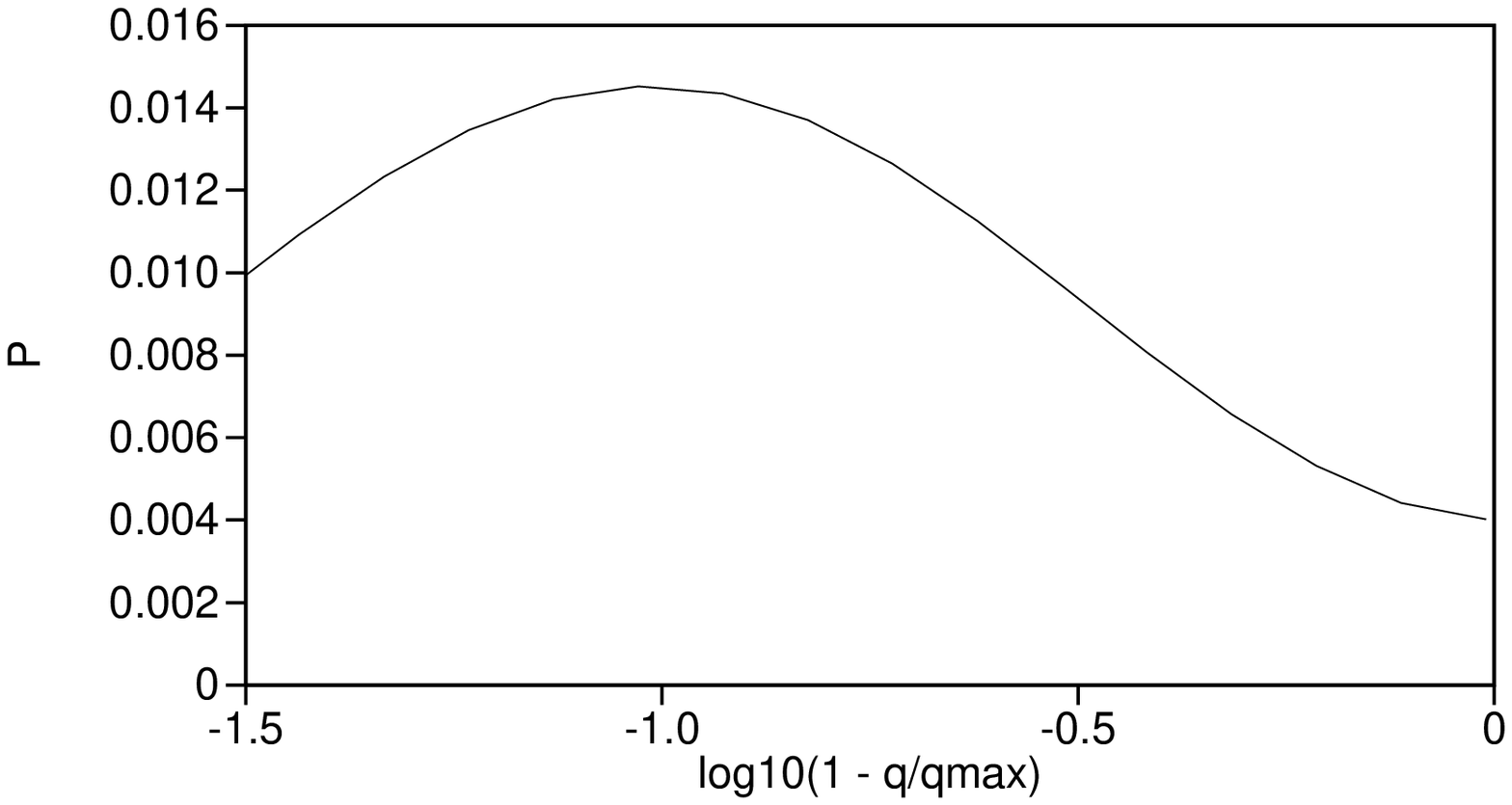}
\vspace*{0.7cm}

{To calculate the rate of production of oppositely charged, {\it
accelerating} black hole pairs in the ${\Bbb Z}_{2}$-symmetric background, we
first restrict the charge to be equal to one of the values $\{q_{n}:\,n
\,\in\, {\Bbb Z}_{+}\}$ such that, for fixed $m$, $\Lambda$, and $\sigma$,
the two periods $\beta_{W}$ and $\beta_{RN}$ satisfy the matching condition}
\[
\beta_{W} = \frac{\beta_{RN}}{n}, ~n \,\in\, {\Bbb Z}_{+}
\]

{As we saw above in Fig. 2, the $q_{n}$ always exist for generic $m$ and
$\Lambda$. We therefore simply need to calculate ${\cal S}_{RN}$ for the
Euclidean section of the accelerating black hole-domain wall configuration.
As above, ${\cal S}_{RN}$ decomposes into a `volume' part and a `domain wall'
part. The calculation is a straightforward generalision of the results of [7]
and yields}
\begin{eqnarray}
\frac{1}{2} {\cal S}_{RN} = 2\pi\sigma \int^{r_{\mbox{\scriptsize max}}}_
{r_{\mbox{\scriptsize min}}}\,dr\,\frac{2\pi\sigma r^{3}}{\sqrt{f\left(f -
(2\pi\sigma r)^{2}\right)}} \nonumber \\
+\, \frac{q_{n}^{2} \beta_{RN}}{4 r_{h}} - \frac{q_{n}^{2}}{2} \int^{r_{\mbox
{\scriptsize max}}}_{r_{\mbox{\scriptsize
min}}}\,dr\,\frac{2\pi\sigma}{f\sqrt{f - (2\pi\sigma r)^{2}}} \\
-\, \frac{\Lambda \beta_{RN} r_{h}^{3}}{12} + \frac{\Lambda}{6} \int^{r_
{\mbox{\scriptsize max}}}_{r_{\mbox{\scriptsize min}}}\,dr\, \frac{2\pi\sigma
r^{4}}{f\sqrt{f - (2\pi\sigma r)^{2}}} \nonumber
\end{eqnarray}

{Thus, the probability that accelerating black hole pairs of mass $m$ and
opposite charge $\pm q_{n}$ will be nucleated in the presence of a ${\Bbb
Z}_{2}$-symmetric wall configuration is given by eq. 27,
where now ${\cal S}_{RN}$ is given by equation (28).

$P$ is again plotted below, for fixed $\Lambda$, $m$ and $\beta$ as $q
\,\longrightarrow\, q_{\mbox{max}}$: Fig. 4.}\\
\vspace*{0.3cm}

\epsfxsize=14cm
\epsfysize=8cm
\psfrag{P}[][]{\raisebox{-0.9cm}{$P$}}
\psfrag{log10(1 - q/qmax)}[][b]{\raisebox{-0.7cm}{$\log_{10}(1 - 
q/q_{\mbox{\scriptsize max}})$}}
\hspace*{-0.7cm} \epsfbox{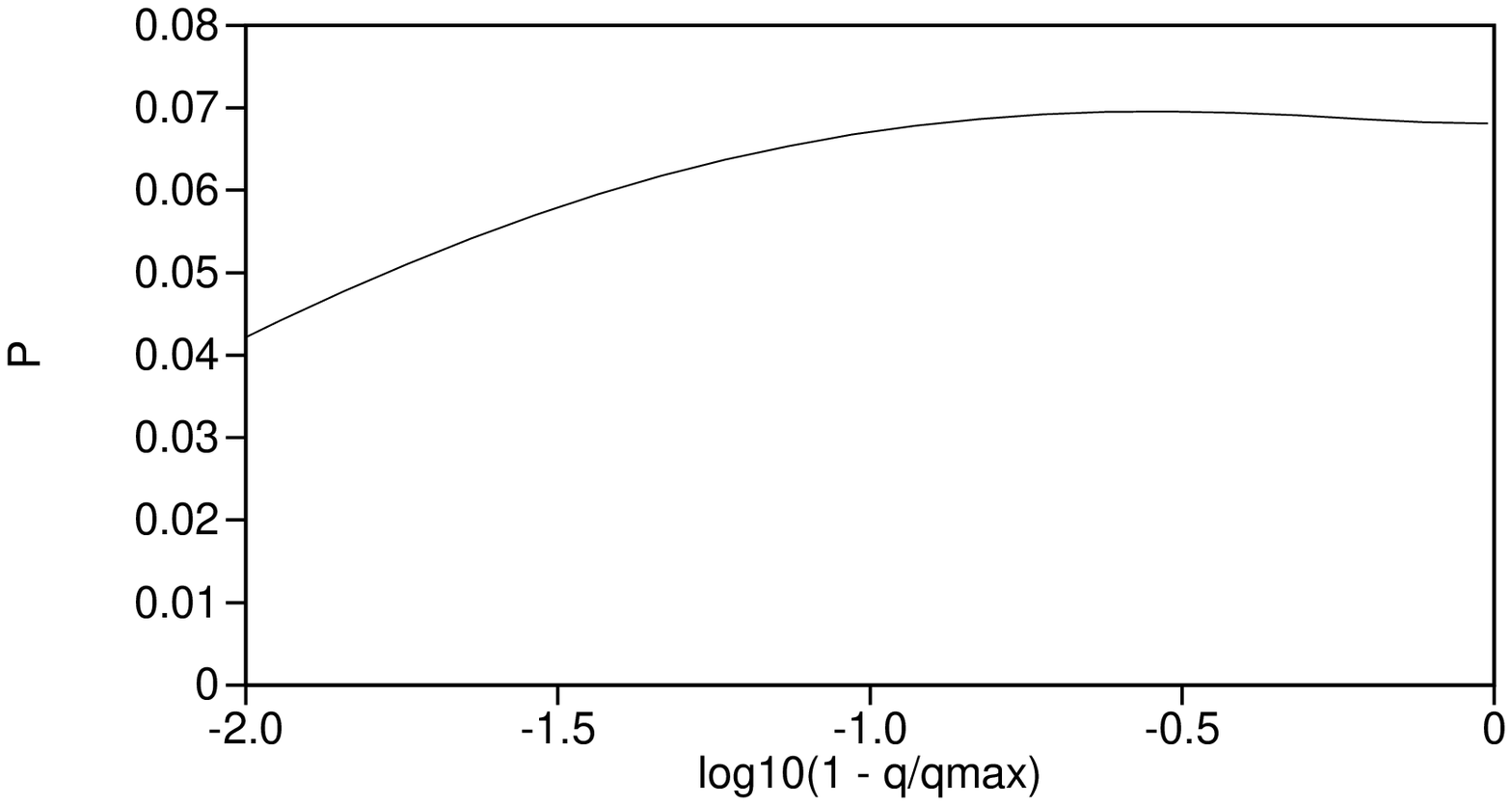}
\vspace*{0.7cm}

{The above calculations were carried out for the values ${\Lambda} = -1$, 
$m = \frac{1}{3\sqrt{3}}$, and
$\beta = 1$; however, the basic properties of the probability distributions 
shown are generic
for all values of these parameters.  In particular, the rate of creation is always 
suppressed as the charge approaches the extremal limit.}\\
\vspace*{0.2cm}

{\noindent \bf Case 2 ($\Lambda_{1} \not= \Lambda_{2}$)}\\

{In the most general situation where the cosmological constant 
varies as we move across the domain wall, we would still expect the 
configuration to
be unstable to a semiclassical process such as black hole pair creation; 
after all, the only
selection rule we need to worry about is charge conservation. The only 
thing exotic
about these non-symmetric walls is that the magnitude of repulsion will 
depend on 
which side you are on. Naively, we would therefore expect that black 
holes of equal
and opposite charge, but unequal mass, would be created in the presence 
of these walls.

However, there is a simple geometrical obstruction to the construction 
of regular
instantons for these configurations; in fact, it would seem that it 
is not possible
to obtain a well behaved Riemannian section for the domain wall, 
much less for the
full black hole-domain wall situation.

We again attempt to construct the instanton for the domain wall by 
taking two 
hyperbolic four-balls, each of radius $\frac{1}{\beta}$, but of 
unequal constant negative
curvatures $\Lambda_{1}$ and $\Lambda_{2}$. We have to be able to join
these two objects along their respective boundaries in order to 
obtain a `nice' instanton
$M_e$ of topology $S^4$ with a ridge of curvature running along 
where the domain wall is.
However, equation (15) for the volume of a three-sphere of radius 
$r$ in hyperbolic
space makes clear that the volumes of the boundaries of the two 
balls will match if and
only if the two cosmological constants are equal! In other words, 
the two balls will be
rather like a large plate and a small bowl; the plate will always 
fit `over' the rim of the
bowl, but it is impossible to align them along their respective 
edges. While one might
devise various clever schemes to get around this pathology, we will 
have nothing more to 
say about these non-symmetric walls in this paper.}\\
\vspace*{0.6cm}

{\noindent \bf 3. No Pair Production of Black Holes by Extreme Vacuum Walls}\\

{As we allow the acceleration parameter $\beta$ to approach zero, 
the size of the wall
tends to infinity until finally, when extremality is reached, the wall 
`decompactifies' and
turns into a static copy of $2+1$ Minkowski space. However, as we have 
seen the 
resulting extreme configuration is still repulsive; hence, one might expect 
the repulsive
energy to provide a source for black hole pair production.

On the other hand, the radius of the instanton obtained in the extremal limit 
is infinite, 
and since the volume contribution to the action (eq. (22)) will always 
dominate 
for large $r$, 
the domain wall action $S_W$ is negative infinite. Thus, in order for the 
process to be
unsuppressed there would have to be a term in the action $S_{RN}$ which 
effectively
counterbalanced the large volume term coming from $S_W$. The only 
candidate term
in $S_{RN}$ which could play such a role is the volume term proportional to 
$\Lambda$. However, this term only grows as $r_{w}^3$, where $r_w$ is the 
`size' of the wall
(recall that this is all consistent since, for very small $\beta$, 
$r_w \sim \frac{1}{\beta}$),
whereas the dominant volume term coming from $S_W$ grows like}
\begin{equation}
\int_{0}^{\frac{1}{\beta}}\left(\frac{1}{\sqrt{|\Lambda|}}\, \sinh
\left(\sqrt{|\Lambda|r}\right)\right)^{3} dr
\end{equation}

{Thus, it is clear that the term which is driving $S_W$ to 
minus infinity will always 
dominate and so the rate will be competely suppressed in the 
extremal limit. Indeed,
one can see this effect numerically; as $\beta$ gets smaller and smaller the 
probability
contours `collapse' down towards zero, until finally the plot of $P$ looks 
effectively like
$P \sim 0$ for all allowed values of the charge. 

Of course, this effect would be expected on the grounds that as one 
approaches the
extremal state one is approaching a BPS state which should be stable 
to all quantum
corrections. This is reminiscent of the situation of black hole pair 
production in the early
universe [19]; there, as one turns down the cosmological constant 
and approaches the stable
Minkowski vacuum state the rate of black hole pair creation is 
totally suppressed in the 
limit. The difference in our example is that the SUSY state which is obtained 
in the limit
is much more complicated and has a richer causal structure than that 
of Minkowski space [1].}\\
\vspace*{0.6cm}

{\noindent \bf 4. Conclusions and Discussion}\\

{We have shown that non-extreme vacuum domain walls which arise in 
$N = 1$, $d = 4$
supergravity are unstable to black hole pair creation; furthermore, 
we found that the
production rate goes to zero as the walls are allowed to approach 
extremality.  This work is an interesting generalization of the work
presented in [7].  The global causal structure of the SUGRA domain
walls studied here is much more subtle than that of the simpler VIS walls
discussed in [7], and furthermore such objects are simple prototypes of
the `braney' type solutions which arise commonly in the supergravity
menagerie.

Of course,
a key ingredient of Kaluza-Klein, supergravity, and effective 
theories derived from string
theory is the dilaton. It is therefore of considerable interest 
to consider the semi-classical
stability of domain walls which are coupled to the dilaton. 
Such dilatonic walls have been
studied for some time, and there is an extensive literature 
about them (see [1] for an 
overview). As would be expected (from the example of dilatonic 
black holes), turning
on the dilaton coupling changes the global spacetime structure 
of the solutions. In fact,
in the non-extreme case these walls often exhibit naked singularities; 
this was one of our
motivations for not considering these walls in this paper. 
Nevertheless, in any regime
where one can find non-extreme configurations which are well-behaved 
we would expect
the basic results of this paper to go through, i.e., 
Reissner-Nordstr{\"o}m-dilaton
anti-de Sitter black holes will be produced by the non-extreme 
backgrounds, and the
rate of production will go to zero as extremality is approached. 
On a related note, it should
also be of interest to see what happens when the walls are 
coupled to gauge fields;
with the introduction of gauge charges it should be possible 
to attain extremality without
resorting to a cosmological constant.

It is worth pointing out that there is a well-defined procedure 
by which these solutions
can be obtained via dimensional reduction of higher dimensional 
configurations in
supergravity (or string theory); in particular, in [20] it was shown 
that the dilatonic
domain wall configurations of Cveti{\v c} et al could be obtained 
by performing 
Scherk-Schwarz dimensional reduction on higher dimensional supergravity 
solutions. The Scherk-Schwarz procedure, which was originally 
introduced to break
SUSY by giving mass to the gravitino, in the present case means 
that some of the 
fields (the axions) are allowed to have a linear dependence on 
a compactification
coordinate. This dependence leads to a cosmological constant 
term in the lower
dimensional, massive supergravity theory. The ($d-2$)-brane 
solutions of the massive
supergravity in four dimensions are precisely the dilatonic 
walls discovered earlier.

Of course, it is well known that black hole solutions can also 
be obtained by taking the 
ordinary Kaluza-Klein reduction of certain higher dimensional 
configurations in string
theory. The fact that both types of object, black hole and domain 
wall, naturally
oxidize to higher dimensional string configurations suggests that 
there might be
an `oxidation' of the semi-classical tunneling process of black 
hole pair creation to 
higher dimensions. Of course, one has to be careful, since at 
least some of the domain
wall energy density (the part proportional to the cosmological 
constant) is coming from
the compactification process. These problems, and many more, 
are currently being actively investigated.}\\
\vspace*{0.6cm}

{\noindent \bf Acknowledgements}\\

{We wish to thank R. Caldwell and R. Emparan for useful discussions. A.C. 
was supported by
NSF PHY94-07194, and J.M.A.A-C was supported by Wolfson College, 
University of Oxford.}\\
\vspace*{0.6cm}

{\noindent \bf References}\\

{\small
{\noindent [1] M. Cveti{\v c} and H.H. Soleng, Phys. Rep. {\bf 282}, 159-223
(1997), preprint hep-th/9604090 (1996).}\\

{\noindent [2] M. Cveti{\v c} and S. Griffies, Phys. Lett. B 285 (1992)
27.}\\

{\noindent [3] M. Cveti{\v c}, S. Griffies and S.J. Rey, Nucl. Phys. 
B 381 (1992) 301.}\\

{\noindent [4] M. Cveti{\v c}, S. Griffies and H.H. Soleng, Phys. Rev. D 
48 (1993) 2613.}\\

{\noindent [5] M. Cveti{\v c}, S. Griffies and H.H. Soleng, Phys. Rev. 
Lett. 71 (1993) 670.}\\

{\noindent [6] M. Cveti{\v c}, S. Griffies and S.J. Rey, Nucl. Phys. 
B 389 (1993) 3.}\\

{\noindent [7] R.R. Caldwell, A. Chamblin and G.W. Gibbons, Phys. Rev. 
D 53 (1996) 7103.}\\

{\noindent [8] G.W. Gibbons, Nucl. Phys. B 395 (1993) 3.}\\

{\noindent [9] G.W. Gibbons, Nucl. Phys. B 472 (1996) 683.}\\

{\noindent [10] M. Cveti{\v c} and S. Griffies, Phys. Lett. B 285 (1992) 
27.}\\

{\noindent [11] R.B. Mann, preprint gr-qc/9607071 (1996).}\\

{\noindent [12] J. Ipser and P. Sikivie, Phys. Rev. D 30 (1984) 712.}\\

{\noindent [13] S.J. Poletti, J. Twamley and D.L. Wiltshire, Phys. Rev. 
D 51 (1995)
5720.}\\

{\noindent [14] J. Louko and S.N. Winters-Hilt, Phys. Rev. D 54 (1996) 
2647.}\\

{\noindent [15] I. Chavel, {\it Riemannian Geometry -- A Modern Introduction},
CUP, Cambridge (1993).}\\

{\noindent [16] M. Cveti{\v c}, preprint hep-th/9306015 (1993).}\\

{\noindent [17] S.W. Hawking and G.F.R. Ellis, {\it The Large-Scale 
Structure of
Spacetime}, CUP, Cambridge (1973).}\\

{\noindent [18] S.W. Hawking and S.F. Ross, Phys. Rev. D 52 (1995) 5865.}\\

{\noindent [19] P. Ginsbarg and M.J. Perry, Nucl. Phys. {\bf B 222}, 245 (1983).}\\

{\noindent [20] P.M. Cowdall, H. Lu, C.N. Pope, K.S. Stelle and P.K. Townsend,
Nucl. Phys. {\bf B 486}, 49-76 (1997).}\\

}

\end{document}